\documentclass[runningheads]{llncs}
\usepackage[T1]{fontenc}
\usepackage{graphicx}
\usepackage{mathtools}
\usepackage{amssymb,amsfonts,stmaryrd}

\usepackage{tikz}
\usetikzlibrary{cd}

\begin{document}
\title{Nonholonomic brackets: Eden revisited\thanks{M.~de León, M.~Lainz and A.~López-Gordón acknowledge financial support from Grants PID2019-106715GB-C21 and CEX2019-000904-S funded by MCIN/AEI/ 10.13039/501100011033. Asier L\'opez-Gord\'on would also like to thank MCIN for the predoctoral contract PRE2020-093814. J.~C.~Marrero ackowledges financial support from the Spanish Ministry of Science and Innovation and European Union (Feder) Grant
PGC2018-098265-B-C32.
}}
%
%
\author{Manuel de León\inst{1,2}\orcidID{0000-0002-8028-2348} \and
Manuel Lainz\inst{1}\orcidID{0000-0002-8028-2348} \and
Asier López-Gordón\inst{1}\orcidID{0000-0002-9620-9647} \and\\
Juan Carlos Marrero\inst{3}\orcidID{0000-0002-9620-9647}}
\authorrunning{M.~de León et al.}
%
\institute{Instituto de Ciencias Matemáticas (CSIC-UAM-UC3M-UCM) \and
Real Academia de Ciencias Exactas, Físicas y Naturales \and
ULL-CSIC Geometría Diferencial y Mecánica Geométrica, Departamento de Matemáticas Estadística e Investigación Operativa and Instituto de Matemáticas y Aplicaciones (IMAULL), University of La Laguna, \protect\\
San Cristóbal de La Laguna, Spain}
\maketitle              
\begin{abstract}
The nonholonomic dynamics can be described by the so-called nonholonomic bracket in the constrained submanifold, which is a non-integrable modification of the Poisson bracket of the ambient space, in this case, of the canonical bracket in the cotangent bundle of the configuration manifold. This bracket was defined in \cite{CdLDMdD,ibort}, although there was already some particular and less direct definition. On the other hand, another bracket, also called noholonomic, was defined using the description of the problem in terms of almost Lie algebroids. Recently, reviewing two older papers by R. J. Eden, we have defined a new bracket which we call Eden bracket. In the present paper, we prove that these three brackets coincide. Moreover, the description of the nonholonomic bracket {\sl \`a la} Eden has allowed us to make important advances in the study of Hamilton-Jacobi theory and the quantization of nonholonomic systems. 

\end{abstract}
	\section{Introduction}
	
	One of the most important objects in mechanics is the Poisson bracket, which allows us to obtain the evolution of an observable by bracketing it with the Hamiltonian energy, or to obtain new conserved quantities of two given ones, using the Jacobi identity satisfied by the bracket. Moreover, the Poisson bracket is fundamental to proceed with the quantization of the system using what Dirac called the analogy principle, also known as the correspondence principle, according to which the Poisson bracket becomes the commutator of the operators associated to the quantized observables.
	
	For a long time, no similar concept has existed in the case of nonholonomic mechanical systems, until van der Schaft and Maschke \cite{vdS} introduced a bracket similar to the canonical Poisson but without the benefit of integrability. 
	In \cite{CdLDMdD,CdLDMdD2}	(see also \cite{ibort}), we have developed a geometric and very simple way to define nonholonomic brackets, in the time-dependent as well time-independent cases. Indeed, it is possible to decompose the tangent bundle and the cotangent bundle along the constraint submanifold in two different ways. Both result in that the nonholonomic dynamics can be obtained by projecting the free dynamics, and furthermore, if we evaluate by the canonical symplectic form the projections of the Hamiltonian fields of two functions on the configuration manifold (prior arbitrary extensions to the whole cotangent), two non-integrable brackets are obtained. The first decomposition is due to de Le\'on and Mart{\'\i}n de Diego \cite{dLDMdD}, and the second one to Bates and Sniatycki \cite{BS}. The advantage of this second decomposition is that it turns out to be symplectic, and it is the one we will use in the present paper. In any case, we proved that both brackets coincided on the submanifold of constraints \cite{CdLDMdD}.
	
	On the other hand, by studying the Hamilton--Jacobi equation, we develop a description of nonholonomic mechanics in the setting of almost Lie algebroids. Note that the ``almost'' is due to the lack of integrability of the distribution determining the constraints, showing the consistency of the description. So, in \cite{dLJCMDMdd} we define a new almost Poisson bracket that we also called nonholonomic. So far, although both nonholonomic brackets have been used in these two different contexts as coinciding, no such proof has ever been published. This paper provides this evidence for the first time. 

But the issue does not end there. In 1951, R.~J.~Eden, who wrote his doctoral thesis on nonholonomic mechanics under the direction of P.~A.~M.~Dirac (see his papers \cite{eden1,eden2}).
In the first paper, Eden introduced an intriguing $\gamma$ operator that mapped free states to constrained states. With that operator (a kind of tensor of type (1,1) that has the properties of a projector), 
Eden obtained the equations of motion, could calculate brackets of all observables, obtained a simple Hamilton--Jacobi equation, and even used it to construct a quantization of the nonholonomic system.
These two papers by Eden have had little impact despite their relevance. Firstly, because they were written in terms of coordinates that made their understanding difficult, and secondly, because it was in the 1980s when 
the study of non-holonomic systems became part of the mainstream of geometric mechanics. 

Recently, we have carefully studied these two papers by Eden, and realized that the operator $\gamma$ is nothing else a projection defined by the orthogonal
decomposition of the cotangent bundle provided by the Riemannian metric given by the kinetic energy. So, we have defined a new bracket that we call Eden bracket, and 
proved that coincides with the previous nonholonomic brackets. 
	
\section{Lagrangian and Hamiltonian mechanics: a brief survey}

\subsection{Lagrangian mechanics}

Let $L : TQ  \longrightarrow \mathbb R$ be a Lagrangian function, 
where $Q$ is a configuration $n$-dimensional manifold. Then, $L = L(q^i, \dot{q}^i)$, where
$(q^i)$ are coordinates in $Q$ and $(q^i, \dot{q}^i)$ are the induced bundle coordinates in $TQ$.
We denote by $\tau_Q : TQ \longrightarrow Q$ the canonical projection such that
$\tau_Q(q^i, \dot{q}^i) = (q^i)$.

We will assume that $L$ is regular, that is, the Hessian matrix $({\partial^2 L}/{\partial \dot{q}^i \partial \dot{q}^j})
$ is regular.
Using the canonical endomorphism $S = d q^i \otimes \frac{\partial}{\partial \dot{q}^i}$ on $TQ$ 
one can construct a 1-form $\lambda_L =  S^* (dL)$, and the 2-form $\omega_L = - d\lambda_L$.
Then, $\omega_L$ is symplectic if and only if $L$ is regular~\cite{dLR}.

Consider now the vector bundle isomorphism $\flat_L  :  T(TQ) \to T^*(TQ)$ given by 
$\flat_L (v) = i_v \, \omega_L$. We define the Hamiltonian vector field $\xi_L = X_{E_L}$ by $\flat_L(\xi_L) = dE_L
$, where $E_L = \Delta(L) -L$ is the energy.
Now, if $(q^i(t), \dot{q}^i(t))$ is an integral curve of $\xi_L$, then it satisfies the 
usual Euler-Lagrange equations
\begin{equation}\label{eleqs}
\frac{d}{dt} \left(\frac{\partial L}{\partial \dot{q}^i}\right) - \frac{\partial L}{\partial q^i} = 0.
\end{equation}

\subsection{Legendre transformation}

Let us recall that the Legendre transformation $FL : TQ
\longrightarrow T^*Q$ is a fibred mapping (that is, $\pi_Q \circ FL
= \tau_Q$, where $\pi_Q : T^*Q
\longrightarrow Q$ denotes the canonical projection of the cotangent bundle of $Q$). Indeed, $FL$ is the fiber derivative of $L$.
In local coordinates, we have
$$
FL (q^i, \dot{q}^i) = (q^i, p_i), \; p_i = \frac{\partial L}{\partial \dot{q}^i},
$$
and thus $L$ is regular if and only if $FL$ is a local diffeomorphism.

Along this paper we will assume that $FL$ is, in fact, a global
diffeomorphism (in other words, $L$ is hyperregular) which is the
case when $L$ is a Lagrangian of mechanical type, say $L=T-V$,
where $T$ is the kinetic energy defined by a Riemannian metric on $Q$, and $V : Q \longrightarrow \mathbb{R}$ is a potential energy.

\subsection{Hamiltonian description}

The Hamiltonian counterpart is developed in the cotangent bundle
$T^*Q$ of $Q$. Denote by $\omega_Q = dq^i \wedge dp_i$ the canonical
symplectic form, where $(q^i, p_i)$ are the canonical coordinates on
$T^*Q$.

The Hamiltonian energy is just $H = E_L \circ (FL)^{-1}$ and the
Hamiltonian vector field is the solution of the symplectic equation
$$
i_{X_H} \, \omega_Q = dH.
$$
As it is well-known, the integral curves $(q^i(t), p_i(t))$ of $X_H$
satisfy the Hamilton equations
\begin{equation}\label{hamiltoneqs}
\dot{q}^i  =  \frac{\partial H}{\partial p_i}, \quad
\dot{p}_i = - \frac{\partial H}{\partial q^i}.
\end{equation}
Since $FL^* \omega_Q = \omega_L$, we deduce that $\xi_L$ and $X_H$
are $FL$-related, and consequently $FL$ transforms the solutions of the
Euler-Lagrange equations \eqref{eleqs} into the solutions of the Hamilton equations \eqref{hamiltoneqs}.

On the other hand, we can define a bracket of functions
$$
\{ \, , \,\}_{can} : C^\infty(T^*Q) \times C^\infty (T^*Q) \longrightarrow C^\infty(T^*Q)
$$
as follows
$$
\{ F , G \}_{can} = \omega_Q(X_ F, X_G) = X_G(F) = - X_F(G)
$$

This bracket is a Poisson bracket, that is, $\{ \, , \,\}_{can} $ is $\mathbb{R}$-bilinear and:
\begin{itemize}
\item it is skew-symmetric: $\{ G , F \}_{can} = - \{ F , G \}_{can}$,
\item it satisfies the Leibniz rule:
$
\{ F F´ , G \}_{can} = F \{ F´ , G \}_{can} + F´\{ F , G \}_{can},
$
and
\item it satisfies the Jacobi identity:
$$
\{ F , \{ G, H\}_{can} \}_{can} + \{ G , \{H, F\}_{can} \}_{can} + \{ H , \{F, G\}_{can} \}_{can} = 0.
$$
\end{itemize}

		
	\section{Nonholonomic mechanical systems}

	\subsection{The Lagrangian description}
	
	A {\bf nonholonomic mechanical system} is a quadruple $(Q, g, V, D)$ where
$Q$ is the configuration manifold of dimension $n$;
	 $g$ is a Riemannian metric on $Q$;
 $V$ is a potential function, $V \in C^\infty(Q)$, and
$D$ is a non-integrable distribution of rank $k < n$ on $Q$.
	
	As in Section 2.2, the metric $g$ and the potential energy $V$ define a Lagrangian function of mechanical type $	L(v_q) = \frac{1}{2} \, g_q(v_q, v_q) - V(q)$.
	
	The nonholonomic dynamics is provided by the Lagrangian $L$ subject to the nonholonomic constraints given by $D$; that means that the permitted velocities should belong to $D$.
	
	The nonholonomic problem is to find the equations of motion
	\begin{equation}
	\frac{d}{dt} \left(\frac{\partial L}{\partial \dot{q}^i}\right) - \frac{\partial L}{\partial q^i} = \lambda_{\alpha} \mu^{\alpha}_{i}(q), \quad \mu^{\alpha}_{i} (q) \dot{q}^i = 0,
	\end{equation}
	where $\{\mu^{\alpha}\}$ is a local basis of $D^\circ$ (the annihilator of $D$) such that
	$\mu^{\alpha} = \mu^{\alpha}_{i} \, dq^i$.
	Here, $\lambda_{\alpha}$ are Lagrange multipliers to be determined.
	
	A geometric description of the above equations can be obtained using the symplectic form $\omega_L$ and the vector bundle of 1-forms, $F$, defined
	by $F = \tau_Q^*(D^\circ)$. So, the above equations are equivalent to these ones
	\begin{eqnarray}
	i_X \, \omega_L - dE_L \in \tau_Q^*(D^\circ), \quad X \in TD.
	\end{eqnarray}
	These equations have a unique solution, $\xi_{nh}$, which is called the nonholonomic vector field.
	
	The Riemannian metric $g$ induces a isomorphism of vector bundles between $TQ$ and $T^*Q$ given by $ \flat_g(q)(v_q) = i_{v_q} g$ (which again induces an isomorphism between vector fields and 1-forms). The inverse of $\flat_g$ will be denoted by $\sharp_g$.
	
	We can define the orthogonal complement, $D^{\perp_g}$, of $D$ with respect to $g$, as follows:
	$$
	D_q^{\perp_g} = \{v_q \in T_qQ \; | \; g(v_q, w_q) = 0, \forall  w_q \in D \}.
	$$
	$D^{\perp_g}$ is again a distribution on $Q$, or, if we prefer, a vector sub-bundle of $TQ$ such that we have the Whitney sum
	\begin{equation}\label{dec1}
	TQ = D \oplus D^{\perp_g}.
	\end{equation}
	
	
	\subsection{The Hamiltonian description}
	
	We can obtain the Hamiltonian description of the nonholonomic system $(Q, g,$ $V, D)$ using the Legendre transformation $FL$, which in our case coincides with the isomorphism $\flat_g$ associated to the metric $g$.

 So, we can define the corresponding
Hamiltonian function $H : T^*Q \longrightarrow \mathbb{R}$, $H = E_L \circ (FL)^{-1} $, and
constraint submanifold 
	$M = FL(D) = \flat_g(D) = (D^{\perp_g})^\circ$.
	Therefore, we obtain a new orthogonal decomposition (or Whitney sum)
	\begin{equation}\label{dec2}
	T^*Q = M \oplus D^\circ,
	\end{equation}
	since $FL (D^{\perp_g}) = \flat_g(D^{\perp_g}) = D^\circ$. This decomposition is orthogonal with respect to the induced metric on tangent covectors, and it is the
	translation of (\ref{dec1}) to the Hamiltonian side.
	Again, $M$ and $D^\circ$ are vector sub-bundles of $\pi_Q : T^*Q \longrightarrow Q$ over $Q$.
	We have the following canonical inclusion $i_M : M \longrightarrow T^*Q$ and orthogonal projection $\gamma : T^*Q \longrightarrow  M$, respectively.

The equations of motion for the nonholonomic system on $T^*Q$ can
now be written as follows:
\begin{equation}\label{hnh}
\dot q^i =\displaystyle{\frac{\partial H}{\partial p_i}}, \quad
\dot p_i = \displaystyle{-\frac{\partial
H}{\partial q^i}- \bar{\lambda}_\alpha \mu^{\alpha}_{j}g^{ij}},
\end{equation}
together with the constraint equations
$\mu^{\alpha}_{i} g^{ij}p_j = 0$
Notice that here the $\bar{\lambda}_{\alpha}$ are Lagrange multipliers to be determined.

Now the vector bundle of constrained forces generated by the 1-forms $\tau_Q^*(\mu^{\alpha})$,
can be translated to the cotangent side and we obtain the vector bundle generated by the 1-forms
$\pi_Q^*(\mu^{\alpha})$, say $\pi_Q^*(D^\circ)$. Therefore, 
the nonholonomic Hamilton equations for  the nonholonomic system can
be then rewritten in intrinsic form as
\begin{equation}\label{a1}
(i_X\omega_Q-dH)_{|M} \in \pi_Q^*(D^\circ), \quad
X_{|M} \in TM.
\end{equation}

These equations have a unique solution, $X_{nh}$, which is called the nonholonomic vector field. Of course, $X_{nh}$ and $\xi_{nh}$ are related by the Legendre transformation restricted to $D$, say, $T(FL)_{|D}(\xi_{nh}) = X_{nh} \circ (FL)_D$.
	
	\subsection{The almost Lie algebroid approach}
	
	In \cite{dLJCMDMdd} (see also \cite{grabowski}) we have developed an approach to nonholonomic mechanics based on the almost Lie algebroid setting.
	
	We denote by $i_D : D \longrightarrow TQ$ the canonical
	inclusion. The canonical projection given by the decomposition $TQ = D \oplus D^\perp$ on $D$
	is denoted by $P : TQ \longrightarrow D$.

	Then, the vector bundle $(\pi_Q)_{|_D} : D \longrightarrow Q$ is an almost Lie algebroid.
	The anchor map is just the canonical inclusion $i_D \longrightarrow TQ$, and the almost Lie bracket $\|\, , \, \|$ on the space of sections $\Gamma(D)$ is given by
	$$
	\|X, Y\| = P([X, Y]),\quad \hbox{for } X, Y \in \Gamma(D).
	$$
	Here, $[\, , \, ]$ is the standard Lie bracket of vector fields.
	
	We also have the vector bundle morphisms provided by the adjoint operators:
	\begin{equation}
	  i_{D}^* : T^*Q \longrightarrow D^*,
	  P^* : D^* \longrightarrow  T^*Q
	\end{equation}
	where  $D^*$ is the dual vector bundle of $D$.
	
	We define now an almost Poisson bracket on $M$ as follows (see \cite{dLJCMDMdd}):
	\begin{eqnarray*}
	&&\{\, , \, \}_{D^*} : C^\infty(D^*) \times  C^\infty(D^*) \longrightarrow C^\infty(D^*)\\
	&&\{f , g\}_{D^*} = \{f \circ i_D^*, g \circ i_D^*\}_{can} \circ P^*.
	\end{eqnarray*}
    The bracket $\{\, , \, \}_{D^*}$ has the same properties as a Poisson bracket except maybe the Jacobi identity, that is, $\{\, , \, \}_{D^*}$ is $\mathbb{R}$-bilinear, skew symmetric and satisfies the Leibniz rule.

    Moreover, if $FL_{nh} : D \longrightarrow D^*$
    is the nonholonomic Legendre transformation given by $FL_{nh} = i_{D^*} \circ FL \circ i_D$ and $Y_{nh}$ is the nonholonomic dynamics in $D^*$, then
    $$
    T(FL_{nh}) (\xi_{nh}) = Y_ {nh} \circ FL_{nh}.
    $$ 
    Hence, the bracket  $\{g , f\}_{D^*}$ may be used to give the evolution of an observable $f \in C^\infty(D^*)$. In fact, if 
    $h : D^* \longrightarrow \mathbb{R}$ is the constrained Hamiltonian function defined by
    $$
    h = (E_L)_{|D} \circ FL_{nh},
    $$
    we have that
    $$
    \dot{f} = Y_{nh}(f) = \{f, h\}_{D^*}, \forall f \in C^\infty(D^*).
    $$

	\section{The nonholonomic bracket}

	Consider the vector sub-bundle $T^DM$ over $M$ defined by
	$$
	T^DM = \{Z \in TM \; | \; T\pi_Q(Z) \in D \}
	$$
	
	As we know \cite{BS,CdLDMdD}, $T^DM$ is a symplectic vector sub-bundle of the symplectic vector bundle 
	$(T_M(T^*Q), \omega_Q)$, where we are denoting again by $\omega_Q$ the restriction to any fiber of $T_M(T^*Q)$.
	Thus, we have the following symplectic decomposition
	\begin{equation}\label{sym2}
	T_M(T^*Q) = T^DM \oplus (T^DM)^{\perp_{\omega_Q}},
	\end{equation}
	where $(T^DM)^{\perp_{\omega_Q}}$ denotes the symplectic complement of $T^DM$.
	Therefore, we have associated projections
	\begin{equation*}
	{\cal P} : T_M(T^*Q) \longrightarrow T^DM, \quad
	{\cal Q} : T_M(T^*Q) \longrightarrow (T^DM)^{\perp_{\omega_Q}}.
	\end{equation*}
    One of the most relevant applications of the above decomposition is that
    \begin{equation*}
        X_{nh} = {\cal P} (X_H), \; \text{along $M$}.
    \end{equation*}
    In addition, the above decomposition allows us to define the so-called nonholonomic bracket as follows.
    Given $f, g \in C^\infty(M)$, we set

\begin{equation}\label{nhb}
\{f, g\}_{nh} = \omega_Q({\cal P}(X_{\tilde{f}}), {\cal P}(X_{\tilde{g}})) \circ i_M
\end{equation}
where $i_M : M \longrightarrow T^*Q$ is the canonical inclusion, and $\tilde{f}, \tilde{g}$ are arbitrary extensions to $T^*Q$ of $f$ and $g$, respectively (see \cite{CdLDMdD,ibort}).
Since the decomposition (\ref{sym2}) is symplectic, one can equivalently write

\begin{equation}\label{nhb2}
\{f, g\}_{nh} = \omega_Q(X_{\tilde{f}}, {\cal P}(X_{\tilde{g}}) ) \circ i_M
\end{equation}

\begin{remark}
Notice that $f\circ \gamma$ and $g\circ \gamma$ are natural extensions of $f$ and $g$ to $T^*Q$, so we can also define the above nonholonomic bracket as follows
\begin{equation}\label{nhb3}
\{f, g\}_{nh} = \omega_Q(X_{f \circ \gamma}, {\cal P}(X_{g \circ \gamma}) ) \circ i_M
\end{equation}
\end{remark}

  It is worth noting that $\{\, , \, \}_{nh}$ is an almost Poisson bracket. Moreover, it may be used to give the evolution of an observable, namely,
    $$\dot f = X_{nh} (f) = \{\, f , H \circ i_M \, \}_{nh}.$$
	
	\section{Eden bracket}
	
	Using the projector $\gamma : T^*Q \longrightarrow M$, we can define an almost Poisson bracket, called the \textbf{Eden bracket}, on $M$ as follows:
	\begin{eqnarray}\label{edenbracket}
	&&\{\, , \, \}_E : C^\infty(M) \times  C^\infty(M) \longrightarrow C^\infty(M) \nonumber\\
	&&\{f , g\}_E = \{f \circ \gamma, g \circ \gamma\}_{can} \circ i_M.
	\end{eqnarray}
    It may be used to give the evolution of an observable, namely,
    $$\dot f = \{\, f , H \circ i_M \, \}_{E}.$$

	\section{Comparison of brackets}

	First of all, one can prove that the almost Poisson brackets defined on $D^*$ and $M$ are isomorphic.

		\begin{theorem}\label{th1}
	The vector bundle isomorphism
	$$
	i_{M, D^*} : M \longrightarrow D^*
	$$
	given by the composition
	$$
	i_{M, D^*}  = i_D^* \circ i_M
	$$
	is an almost Poisson isomorphism between the almost Poisson manifolds
	$(M, \{\, , \, \}_{E})$ and  $(D^*, \{\, , \, \}_{D^*})$ .

	\end{theorem}

	Additionally, one can prove that the Eden bracket is just the nonholonomic bracket.
	
	\begin{proposition}
	We have
 \begin{itemize}
     \item ${\cal P} (Z) = T\gamma(Z)$ for every $Z \in T^DM$.
     \item 	For any function $f \in C^{\infty}(M)$, we have 
     $T\pi_Q(x)(X_{f \circ \gamma}) \in D_{\pi_Q(x)}$
	for every $x \in M$. In consequence, $X_{f \circ \gamma} (x) \in (T^DM)_x, \forall x \in M$.

 \end{itemize}	
	\end{proposition}

	\begin{theorem}
	The identity map
	$$
	\operatorname{id}\colon (M, \{\, , \, \}_{nh}) \longrightarrow (M, \{\, , \, \}_{E}) 
	$$
	is an almost Poisson isomorphism.
	\end{theorem}

\end{document}